%% file: lat98_revision1.tex
\documentstyle[twoside,fleqn,espcrc2,epsf]{article}

\input mydefs

\newcommand{\AmS}{{\protect\the\textfont2
  A\kern-.1667em\lower.5ex\hbox{M}\kern-.125emS}}
\newcommand{\fB}{$f_B$}
\newcommand{\fBs}{$f_{B_s}$}
\newcommand{\fD}{$f_D$}
\newcommand{\fDs}{$f_{D_s}$}
\newcommand{\fBsofB}{$f_{B_s}/f_B$}
\newcommand{\fDsofD}{$f_{D_s}/f_D$}

\title{Heavy-Light Decay Constants:  Conclusions from the Wilson Action}

\author{ C.~Bernard,\hskip-0.03in
\address{{\vskip-0.10in{\hskip 0.07in Department of Physics, Washington
University, St.~Louis, MO 63130, USA}}} 
\thanks{presented by C.~Bernard}
T.~DeGrand,\hskip-0.03in
\address{Physics Department, University of Colorado, Boulder, CO 80309, USA} %
C.~DeTar,\hskip-0.03in
\address{Physics Department, University of Utah, Salt Lake City, UT 84112, USA}
Steven~Gottlieb,\hskip-0.03in
\address{Department of Physics, Indiana University, Bloomington, IN 47405, USA}
Urs~M.~Heller,\hskip-0.03in
\address{SCRI, Florida State University, Tallahassee, FL 32306-4130, USA} 
J.~Hetrick,\hskip-0.03in
\address{Department of
Physics, University of the Pacific, Stockton, CA 95211, USA} 
N.~Ishizuka,\hskip-0.03in
\address{Institute of Physics, University of Tsukuba, Tsukuba, Ibaraki 305, Japan} 
C.~McNeile,\hskip-0.03in$\,\null^{\rm c}$
R.~Sugar,\hskip-0.03in
\address{Department of Physics, University of California, Santa Barbara, CA
93106, USA} 
D.~Toussaint,\hskip-0.03in
\address{Department of Physics, University of Arizona, Tucson, AZ 85721, USA} 
and M.~Wingate\hskip-0.03in
\address{RIKEN BNL Research Center, Upton, New York 11973, USA} 
} 

\begin{document}

\begin{abstract}
We report on the results of a MILC collaboration calculation
of $f_B$, $f_{B_s}$, $f_D$, $f_{D_s}$ and their ratios.  We discuss
the most important errors in more detail than we have elsewhere.

\end{abstract}

\maketitle

As is well known, precise computations of heavy-light decay
constants such as \fB\ would place stringent constraints on the
Standard Model.  
Reference~\cite{MILC-PRL} describes our recently completed evaluation 
of  these decay constant using the Wilson action.
We have good control of all sources of
error within the quenched approximation.  By comparing our
quenched results with those from lattices with two flavors of
staggered dynamical quarks, we are also able to estimate
the error due to quenching.  However, for reasons
described below, the error on this error is probably rather large.

Rather than repeat the full exposition of Ref.~\cite{MILC-PRL},
we concentrate only on a few key points here and discuss them
in greater detail than was possible previously. 
This paper should
therefore be read in conjunction with \cite{MILC-PRL}, which also
contains references to related work.
Further, to complement the earlier discussion, we focus on our data for
the ratios of decay constants (\fBsofB\ and \fDsofD), rather than
the decay constants themselves.

\begin{figure}[htb] 
\vspace{-44pt}
\epsfxsize=1.0 \hsize
\epsffile{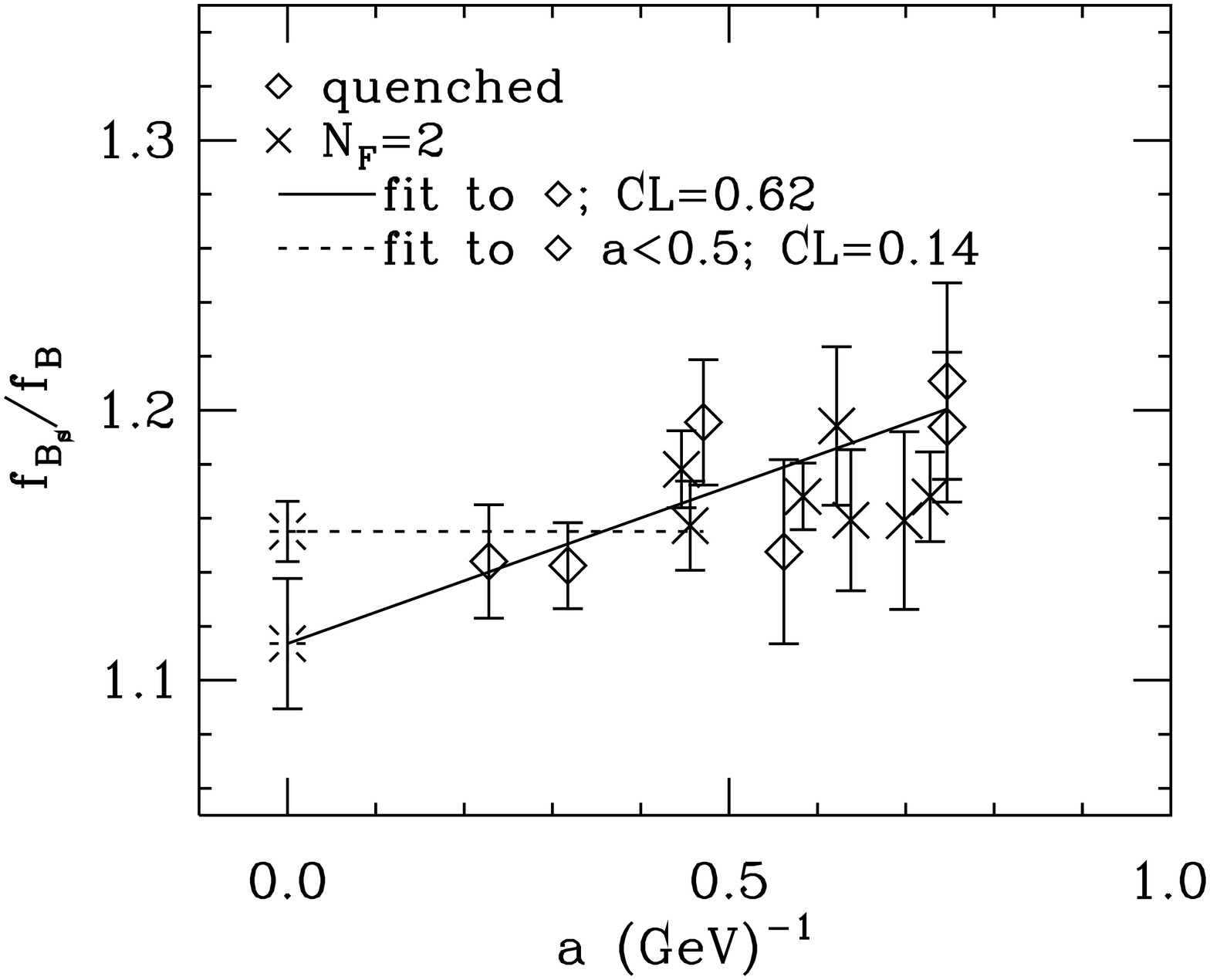}
\vspace{-28pt}
\caption{ Results for \fBsofB\ as a function of lattice spacing,
with central choices for other elements of the analysis.
The fits are to the
quenched approximation points (diamonds) only.
}
\vspace{-18pt}
\label{fig:fbsofb}
\end{figure}

\begin{figure}[htb] 
\vspace{-44pt}
\epsfxsize=1.0 \hsize
\epsffile{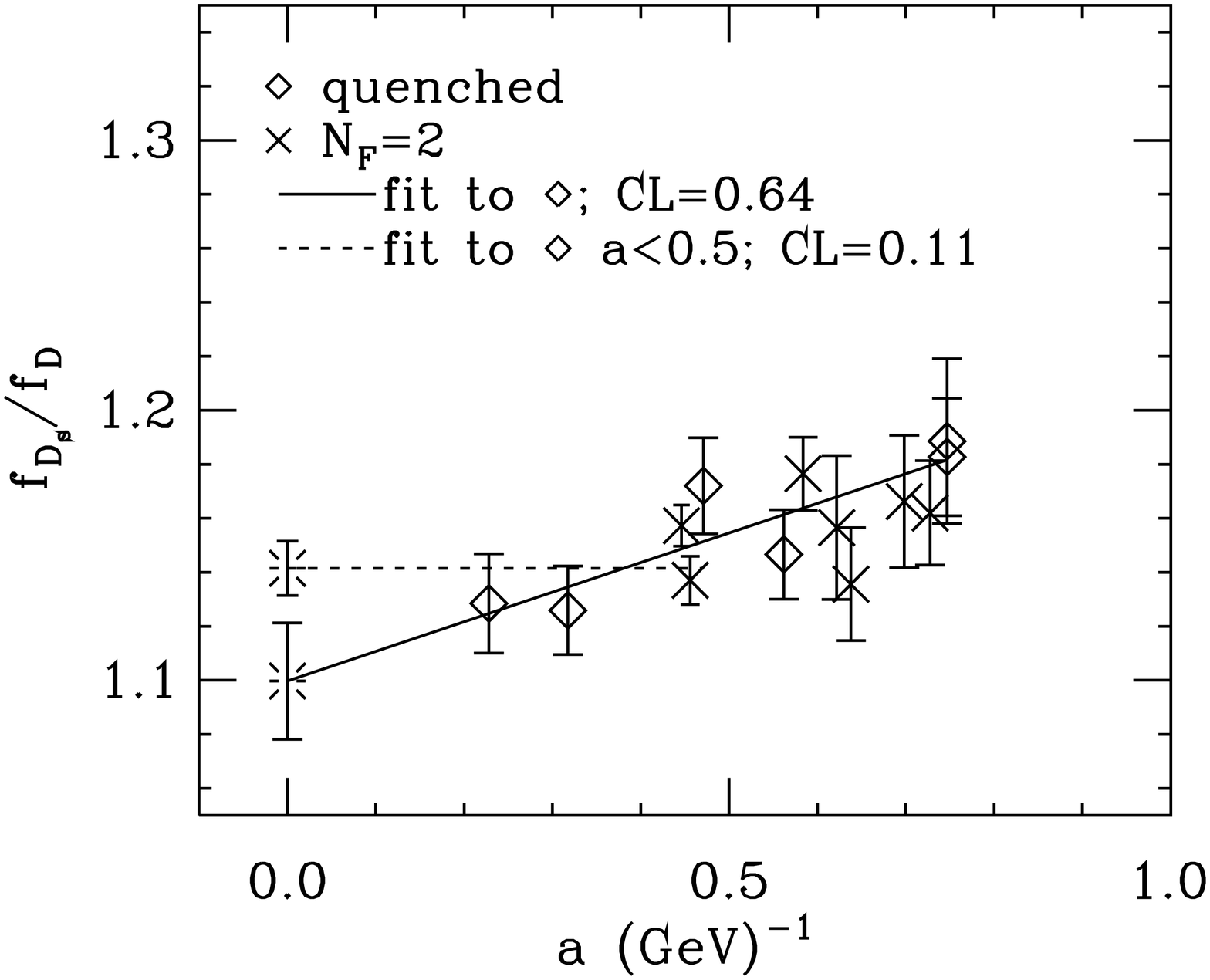}
\vspace{-28pt}
\caption{}{ Same as Fig.~\ref{fig:fbsofb},
but for \fDsofD.}
\vspace{-18pt}
\label{fig:fdsofd}
\end{figure}

The largest error within the quenched approximation comes
from (a) the extrapolation to the continuum, in conjunction with
(b) the chiral extrapolation, and (c) perturbation theory errors.
(Errors (a), (b), and (c) cannot be disentanged 
with current data.)  Figures~\ref{fig:fbsofb} and \ref{fig:fdsofd}
show \fBsofB\ and \fDsofD, respectively, as 
functions of lattice spacing.
In each case, 
two possible extrapolations to the continuum are shown:  (1) a linear
fit in $a$ to all the quenched data, and (2) a constant
fit to all the quenched data with $a\ltwid0.5\ \GeV^{-1}$
($6/g^2 \ge 6.0$).  Since we expect
Wilson fermions to have $\cO(a)$ errors, fit (1) is the most natural
choice.  This choice receives support from the confidence
levels of the fits: in both figures, (1) is a much better
fit than (2).

For the decay constants themselves, unlike the ratios, the
confidence levels of the linear and constant fits described
above are both good, with the constant fits in fact 
having higher confidence levels than the linear
fits. (See Fig.~1 in \cite{MILC-PRL}.)  However it would
be inconsistent to treat the decay constants as independent
of $a$ (for $a\ltwid0.5\ \GeV^{-1}$), yet fit the ratios linearly.
We therefore choose, for our central values of both 
decay constants and ratios, the results
from the linear fit.
Clearly, however, the difference between the fits must be
included in the systematic error estimate.

As described in Ref.~\cite{MILC-PRL}, the errors coming
from the chiral extrapolations, as well as perturbative effects
beyond one loop, are entangled with the continuum extrapolation
error.  This is because a change in the types of chiral fits used,
or a change in the one-loop scale $q^*$, moves the  individual,
fixed-$a$ points enough to affect significantly the difference
between linear and constant continuum extrapolations.  

Altogether, we consider 4 choices for the chiral fits and 3 choices
for $q^*$
(see \cite{MILC-PRL}).  
We then compute each quantity 24 times (2 continuum extrapolations $\times$
4 chiral fits $\times$ 3 scale choices), giving a central
value and  23 alternatives.  The spread in the alternatives 
(taken separately in the positive and negative directions)
determines the combined systematic error due to these three effects.

For the decay constants, the errors due to each of the
three effects alone (determined by variation in the corresponding choices only)
are comparable, and the combined error computation is nontrivial.
For example, 
the combined positive error
thus determined on $f_B$ or $f_{B_s}$ is
$\sim\!25\%$ smaller than the sum in quadrature of the 
three individual errors.
This is due to the correlations among the errors.  

\begin{figure}[htb] 
\vspace{-44pt}
\epsfxsize=1.0 \hsize
\epsffile{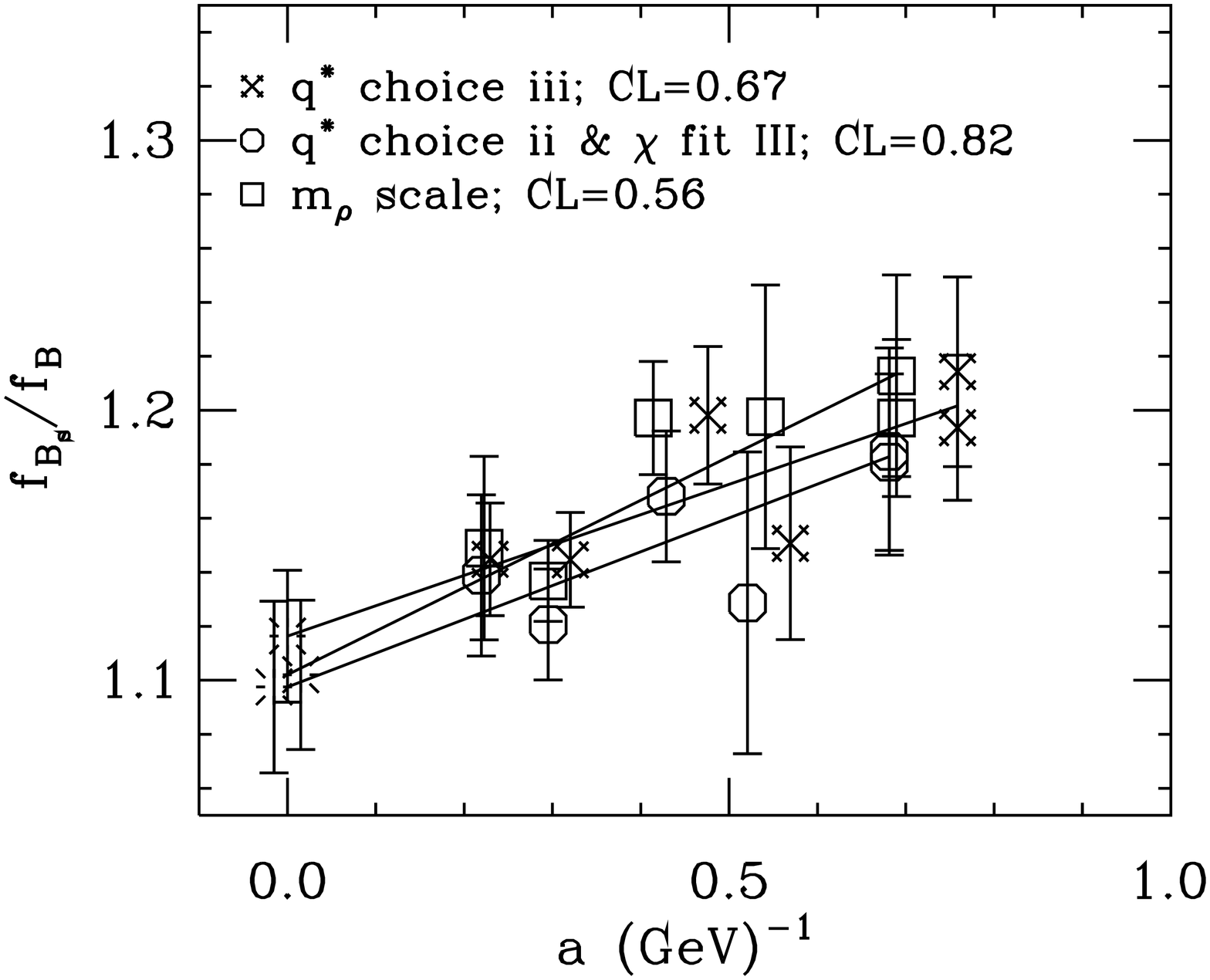}
\vspace{-28pt}
\caption{}{ Quenched \fBsofB\ with various alternative choices in the
analysis.}
\vspace{-18pt}
\label{fig:fbsofb_alt}
\end{figure}

In contrast, for the ratios of decay constants,
the errors due to the chiral fits or perturbative corrections alone
are quite small.  This is illustrated for
\fBsofB\ in Fig.~\ref{fig:fbsofb_alt}.
Changing the chiral fit and/or the $q^*$ choice 
makes very little
difference, as long as the linear continuum extrapolation is used
in all cases.  The combined error for each of the ratios
then turns out to be, to a good approximation,
simply the average (over all chiral
and $q^*$ choices) of the difference between the linear and
constant continuum extrapolations.

Other sources of error within the quenched approximation 
are much smaller than the combined error just described
and appear to be more
or less independent of it. They
are added to the combined error
in quadrature.  See Ref.~\cite{MILC-PRL} for details.

We emphasize that the central values quoted come from the quenched
approximation.  Our $N_F\!=\!2$ dynamical fermion simulations
are used only to make an estimate of the effects of including
virtual quark loops.   We feel it would be premature to try
quote ``full QCD'' results because: (1) the virtual quark mass
is fixed and not extrapolated to the chiral limit,
(2) we do not believe the $N_F\!=\!2$
data is 
good enough at this point to
attempt an extrapolation to $a\!=\!0$, and (3) we have two light flavors,
not three. 

In practice, the quenching error is estimated primarily by comparing the 
$N_F\!=\!2$ results at the smallest available lattice spacing 
($a\approx 0.45\ {\rm  GeV}^{-1}$, $\beta=5.6$, $am=0.01$, $16^3\times 32$ and
$24^3\times 64$) with the quenched results interpolated ({\it via} the
linear fit) to the same lattice spacing.  The $24^3\times 64$ $N_F\!=\!2$
set is new this year and gives decay constants which are consistent
with, but somewhat larger than, the decay constants on the older
 $16^3\times 32$ (HEMCGC) set.
Since the $N_F\!=\!2$ decay constants at this lattice
spacing are always larger than the corresponding quenched
ones (see Fig.~1 of Ref.~\cite{MILC-PRL}), we end up quoting a
very asymmetric quenching error for \fB, \fBs, \fD, and \fDs.  
For \fBsofB\ and \fDsofD,
the results for $N_F\!=\!2$ and quenched
lattices are consistent within errors (see \eg Fig.~\ref{fig:fbsofb}).  
It is not surprising therefore that the
quenching error we estimate in this case is therefore roughly the
size of the 
statistical errors.  

The effect of changing the way the lattice spacing is determined (from fixing
$f_\pi$ to fixing $m_\rho$) can also be used as a rough estimate
of the size of (some) quenching errors.  For the decay constants,
this method  gives an error much smaller than the direct comparison
described above.  For \fBsofB\ and \fDsofD, the two approaches give
comparable estimates (compare Fig.~\ref{fig:fbsofb_alt} with
Fig.~\ref{fig:fbsofb}).  Similar
statements apply to the effect
of changing the way the strange quark hopping parameter is determined
(from fixing the $K$ mass to fixing the $\phi$ mass).
We take the final quenching error as the largest error found with
any of the methods.

Our results are (in MeV):
\vspace{-4pt}
\begin{eqnarray*}
f_B \!=\! 157 (11) ({}^{+25}_{-9})(  {}^{+23}_{ -0});\
f_{B_s}\!\!\!\! \!&=&\!\!\!\!\! 171 (10) ({}^{+34}_{ -9})(  {}^{+27}_{ -2})
\\
f_D \!=\! 192(11) ({}^{+16}_{ -8} )({}^{+15}_{ -0});\ \
\!f_{D_s} \!\!\!\!\!&=&\!\!\!\!\! 210 (9) ({}^{+25}_{ -9} )({}^{+17}_{ -1}).
\vspace{-4pt}
\end{eqnarray*}
The errors are statistical,
systematic (within the quenched
approximation), and systematic (of quenching), respectively.
It must be kept in mind
that the error on the quenching error is large.  
For example,
an extrapolation of the $N_F\!=\!2$ results
to the continuum looks like it would significantly
increase the (positive) quenching error on the decay constants.
Further, even within the current algorithm for estimating
quenching, the error on the quenching error is $\gtwid 50\%$,
due to the statistical fluctuations and the various systematic
variations in the analysis. 

For the ratios, we find:
\vspace{-1pt}
\begin{eqnarray*}
f_{B_s}/f_B \!\!&=&\!\! 1.11 (2) ({}^{+4}_{ -3})  (3)\quad
\phantom{f_{B_s} = 164 (9) }\\
f_{D_s}/f_D\!\! &=&\!\! 1.10 (2) ({}^{+4}_{ -2})  ({}^{+2}_{ -3})\quad
\phantom{f_{B_s} = 164 (9) }\\
f_{B}/f_{D_s}\!\! &=&\!\! 0.75 (3)({}^{+4}_{ -2})({}^{+8}_{ -0})\quad
\phantom{f_{B_s} = 164 (9) }\\
f_{B_s}/f_{D_s} \!\! &=&\!\! 0.85 (3) ( {}^{+5}_{ -3}) ({}^{+5}_{ -0}).\quad
\phantom{f_{B_s} = 164 (9) }
\vspace{-8pt}
\end{eqnarray*}

We note that as experimental measurements of \fDs\ 
improve, the ratios $f_{B}/f_{D_s}$
and $f_{B_s}/f_{D_s}$ may ultimately provide the best way to determine \fB\ and
\fBs.  

We thank Y.\ Kuramashi for
disucssions and CCS (ORNL), SDSC,
Indiana University, NCSA,
PSC, MHPCC, CTC, CHPC (Utah), and Sandia Natl.\ Lab.\
for computing resources.
This work was supported in part by the DOE and NSF.

\end{document}

%% file: mydefs.tex
\def\eg{{\it e.g.},\ }
\def\et{{\it et al.}}

\def\gtwid{\raise.3ex\hbox{$>$\kern-.75em\lower1ex\hbox{$\sim$}}}
\def\ltwid{\raise.3ex\hbox{$<$\kern-.75em\lower1ex\hbox{$\sim$}}}

\def\eg{{\it e.g.},\ }
\def\et{{\it et al.}}

\def\cO{{\cal O}}

\def\nsection#1 #2{\leftline{\rlap{#1}\indent\relax #2}}

\def\GeV{{\rm Ge\!V}}

%% file: lat98_revision1.bbl
\begin{thebibliography}{9}
\vspace{-4pt}

\bibitem{MILC-PRL}  C.\ Bernard, \et,
hep-ph/9806412, submitted to Phys.\ Rev.\ Lett.

\end{thebibliography}
